\documentclass{article}
\usepackage{amsmath, amsfonts, amssymb, amsthm}
\usepackage{graphicx}
\usepackage{here}
\begin{document}

\begin{center}
{\large\bf
Neutrino oscillation phenomenology and
  impact of Professor Masatoshi Koshiba}
\end{center}
\vspace*{.6cm}

\begin{center}
\large{\sc Osamu Yasuda}
\end{center}
\vspace*{0cm}
{\it
\begin{center}
Department of Physics, Tokyo Metropolitan University,\\
Hachioji, Tokyo 192-0397, Japan
\end{center}}

\vspace*{0.5cm}

{\Large \bf
\begin{center} Abstract \end{center}  }
Neutrino oscillation
phenomenology is briefly reviewed,
and impact of the late Professor Masatoshi Koshiba
on research on neutrino oscillation
is discussed from the viewpoint of phenomenology.
\vspace*{0.5cm}

\section{Introduction}
Since the announcement of the discovery of atmospheric
neutrino oscillation by the Superkamiokande Collaboration\,\cite{Fukuda:1998mi}
in 1998, there has been remarkable progress in research on
neutrino oscillation.  In this article I briefly describe
neutrino oscillation phenomenology
and I emphasize the impact of Professor Koshiba
on neutrino oscillation study.

\section{Three flavor neutrino oscillation}
\subsection{Preliminary}
If we assume neutrino mass, we have to describe
neutrinos in terms of the Dirac equation
for spinors with masses.
The mass eigenstates $\nu_j~(j=1,2,3)$ of neutrinos
with mass $m_j$ in vacuum
can be described by the free Dirac equation.
The Dirac equation of a spinor with mass $m$ and
momentum $\vec{p}$
has the energy eigenvalues $E,E,-E,-E$
where $E\equiv\sqrt{\vec{p}^{\,2}+ m^2}$.
If we extract one component with the positive energy of $\nu_j$,
then the mass eigenstates $\nu_j~(j=1,2,3)$ of neutrinos
satisfy the Dirac equation in vacuum
  \begin{eqnarray}
&{\ }& \hspace{-41mm}
  \displaystyle i {d {\ }\over dt} \nu_j(t)=
E_j~\nu_j(t)\,,\quad E_j\equiv\sqrt{\vec{p}^{\,2}+ m_j^2}.
\label{sch1}
\end{eqnarray}

On the other hand, the flavor eigenstates of neutrinos
($(\nu_e,\nu_\mu,\nu_\tau)^T$)
are those which can be observed
in the processes
$\nu_e + n \to e^- + p$,
$\nu_\mu + n \to \mu^- + p$,
$\nu_\tau + n \to \tau^- + p$.
The mass ($(\nu_1,\nu_2,\nu_3)^T$) and
flavor ($(\nu_e,\nu_\mu,\nu_\tau)^T$) eigenstates
of the neutrino sector are related by
\begin{eqnarray}
&{\ }& \hspace{-61mm}
  \left( \begin{array}{c}
    \nu_e\\
    \nu_\mu\\
    \nu_\tau
    \end{array}
    \right)
= U
  \left( \begin{array}{c}
    \nu_1\\
    \nu_2\\
    \nu_3
    \end{array}
    \right)\,,
    \nonumber
\end{eqnarray}
where
\begin{eqnarray}
&{\ }& \hspace{-10mm}
  U\equiv
\left(
\begin{array}{ccc}
 1 &0 &0 \\
0  &c_{23} &s_{23} \\
0  &-s_{23}  &c_{23}  \\
\end{array}
\right)
\left(
\begin{array}{ccc}
c_{13}  &0 &s_{13}e^{-i\delta} \\
0  &1 &0 \\
-s_{13}e^{i\delta}  &0 &c_{13} \\
\end{array}
\right)
\left(
\begin{array}{ccc}
c_{12}  &s_{12} &0 \\
-s_{12}  &c_{12} &0 \\
0  &0 &1 \\
\end{array}
\right)
  \nonumber\\
&{\ }& \hspace{-6mm}
=\left(
\begin{array}{ccc}
c_{12}c_{13} & s_{12}c_{13} &  s_{13}e^{-i\delta}\\
-s_{12}c_{23}-c_{12}s_{23}s_{13}e^{i\delta} &
c_{12}c_{23}-s_{12}s_{23}s_{13}e^{i\delta} & s_{23}c_{13}\\
s_{12}s_{23}-c_{12}c_{23}s_{13}e^{i\delta} &
-c_{12}s_{23}-s_{12}c_{23}s_{13}e^{i\delta} & c_{23}c_{13}\\
\end{array}
\right)
\label{mns}
\end{eqnarray}
is the $3\times 3$ unitary neutrino mixing
matrix\,\cite{Pontecorvo:1957cp,Maki:1962mu},
$\theta_{12}$, $\theta_{23}$, $\theta_{13}$
are three mixing angles,
$\delta$ is a CP phase,
and $c_{jk}\equiv\cos\theta_{jk}$, $s_{jk}\equiv\sin\theta_{jk}$.
It is known that the propagation of neutrinos
in matter is described with the matter effect
\cite{Wolfenstein:1977ue,Mikheyev:1985zog}:
\begin{eqnarray}
&{\ }& \hspace{-3mm}
\displaystyle i {d \over dt} \left( \begin{array}{c} \nu_e(t) \\
\nu_\mu(t)\\\nu_\tau(t)
\end{array} \right)
=M~\left( \begin{array}{c} \nu_e(t) \\
\nu_\mu(t)\\\nu_\tau(t)
\end{array} \right)\qquad\qquad\qquad\qquad
\label{sch3}\\
&{\ }& \hspace{-3mm}
M\equiv[U\,\mbox{\rm diag}(E_1,E_2,E_3)\,U^{-1}
  +\mbox{\rm diag}(A_{\mbox{\tiny CC}}+A_{\mbox{\tiny NC}},A_{\mbox{\tiny NC}},A_{\mbox{\tiny NC}})
  ]
\nonumber\\
&{\ }& \hspace{3mm}
=[U\,\mbox{\rm diag}(0,\Delta E_{21},\Delta E_{31})\,U^{-1}
  +\mbox{\rm diag}(A_{\mbox{\tiny CC}},0,0)+(E_1+A_{\mbox{\tiny NC}})\,{\bf 1}]\,,
\label{m3}
\end{eqnarray}
where $\Delta E_{jk}\equiv E_j - E_k\simeq (m_j^2 - m_k^2)/2|\vec{p}|
\equiv\Delta m_{jk}^2/2|\vec{p}|\simeq \Delta m_{jk}^2/2E$,
\begin{eqnarray}
&{\ }& \hspace{-0mm}
  A_{\mbox{\tiny CC}} \equiv \sqrt{2} \,G_F N_e
  = \left[\frac{\rho}
    {2.6 ~\mathrm{(g \cdot cm^{-3})}}\right]
  \cdot \left(\frac{Y_e}{0.5}\right)
    \cdot 1.0 \times 10^{-13} \,\mbox{\rm eV}
\nonumber\\
&{\ }& \hspace{-0mm}
A_{\mbox{\tiny NC}} \equiv  -\frac{1}{\sqrt{2}} \,G_F N_n
= -\left[\frac{\rho}
    {2.6 ~\mathrm{(g \cdot cm^{-3})}}\right]
  \cdot \left(\frac{1-Y_e}{0.5}\right)
    \cdot 5.0 \times 10^{-14} \,\mbox{\rm eV}
\nonumber
\end{eqnarray}
stand for
the matter effect due to
the charged and neutral current interactions,
$N_e$ and $N_n$ stand for the density
of electrons and neutrons, $G_F= 1.166\times 10^{-5}$ GeV$^{-2}$ is
the Fermi coupling constant, and $Y_e=N_p/(N_p+N_n)$ is the
relative number density of electrons in matter,
where $N_p$ represents the number of protons
per unit volume.
Since the term which is proportional to the $3\times 3$ identity matrix ${\bf 1}$
only affects the over all phase of the probability
amplitude $A(\nu_\alpha \to \nu_\beta)$,
$(E_1+A_{\mbox{\tiny NC}})\,{\bf 1}$ in Eq.\,(\ref{m3})
can be ignored.

\subsection{Mixing angles and mass squared differences}
If the density is constant, then Eq.\,(\ref{m3}) can be
formally diagonalized as
\begin{eqnarray}
&{\ }& \hspace{-61mm}
M
=\tilde{U}\,{\mbox{\rm diag}}\left(
\tilde{E}_1,\tilde{E}_2,\tilde{E}_3\right)
\,\tilde{U}^{-1},
\nonumber
\end{eqnarray}
where $\tilde{E}_j~(j=1,2,3)$ are
the energy eigenvalues of $M$ in matter, and
$\tilde{U}$ is a unitary matrix which diagonalizes $M$.
The probability $P(\nu_\alpha\rightarrow\nu_\beta)$
can be expressed as
\begin{eqnarray}
&{\ }& \hspace{-10mm}
P(\nu_\alpha\rightarrow\nu_\beta)=\delta_{\alpha\beta} -
4\sum_{j<k}\mbox{\rm Re}(\tilde{U}_{\alpha j}\tilde{U}_{\beta
j}^\ast \tilde{U}_{\alpha k}^\ast \tilde{U}_{\beta
k})\sin^2\left( \frac{\Delta \tilde{E}_{jk}L}{2}\right)
\nonumber\\ 
&{\ }& \hspace{15mm}
-2\sum_{j<k}\mbox{\rm
Im}(\tilde{U}_{\alpha j}\tilde{U}_{\beta j}^\ast
\tilde{U}_{\alpha k}^\ast \tilde{U}_{\beta k})\sin( \Delta
\tilde{E}_{jk}L)\,,
\label{probm}
\end{eqnarray}
where $\Delta \tilde{E}_{jk}\equiv \tilde{E}_j-\tilde{E}_k$.

From the parametrization (\ref{mns}),
the simplest mixing angle to determine
is $\theta_{13}$.  If we consider the
disappearance experiment of reactor neutrinos
with a baseline length $L$=2 km,
where the mean neutrino energy is $E\simeq$ 4 MeV,
and if we assume that the contribution from
the smaller mass squared difference
$\Delta m_{21}^2 L/2E$ is small,
then, because the matter effect, which
appears in the form of $A_{\mbox{\tiny CC}}L/2\sim
L/(4000~\mbox{\rm km})$ in the
oscillation probability, can be ignored,
and we obtain
\begin{eqnarray}
&{\ }& \hspace{-20mm}
P(\bar{\nu}_e \rightarrow \bar{\nu}_e) 
\simeq 1-4|U_{e3}|^2(1-|U_{e3}|^2)
\sin^2\left(\frac{\Delta m_{31}^2 L}{4E}\right)
\nonumber\\ 
&{\ }& \hspace{1mm}
= 1-\sin^22\theta_{13}
\sin^2\left(\frac{\Delta m_{31}^2 L}{4E}\right)\,.
\label{chooz}
\end{eqnarray}
Eq.\,(\ref{chooz}) has the same form as the oscillation
probability in the two flavor framework, and
one can get information on $\theta_{13}$
from the result which is expressed in terms of the two flavor
oscillation analysis.
From the negative result of the CHOOZ reactor neutrino
experiment\,\cite{CHOOZ:1999hei},
we obtain
\begin{eqnarray}
&{\ }& \hspace{-10mm}
  \sin^22\theta_{13} \lesssim 0.15
  \quad\mbox{\rm for}~|\Delta m_{31}^2| = 2.5\times 10^{-3}\mbox{\rm eV}^2
  ~(90\%~\mbox{\rm CL})
\label{theta13}
\end{eqnarray}

\begin{figure}[H]
\begin{center}
\hglue -1.4cm
\includegraphics[scale=0.3]{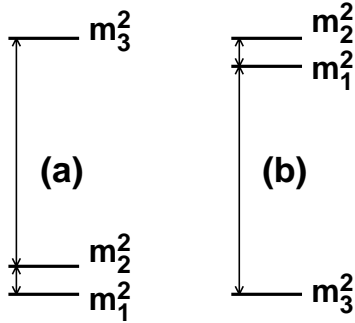}
\hglue -0.4cm
\vglue -0.4cm
\caption{Two mass patterns:
(a) Normal Ordering ($\Delta m^2_{31}>0$), 
  (b) Inverted Ordering ($\Delta m^2_{31}<0$).
  Both mass patterns are allowed as of 2021.
}
\label{fig1}
\end{center}
\end{figure}

Eq.\,(\ref{theta13}) shows that $\theta_{13}$ is small,
so we can put $\theta_{13}\to 0$ in the zeroth approximation.
If we consider the case where
$|\Delta E_{21}|\simeq|\Delta m_{21}^2 /2E|\ll |\Delta E_{31}|
\simeq |\Delta m_{31}^2 /2E| \sim A_{\mbox{\tiny CC}}$,
then the term which is proportional to $\Delta E_{21}$
in Eq.\,(\ref{m3}) can be ignored, so $\theta_{12}$
as well as $\theta_{13}$ disappear, the $\nu_e$ state
decouples from the $\nu_\mu$ and $\nu_\tau$ states
(i.e., the matter effect disappears from
the sector of $\nu_\mu$ and $\nu_\tau$), and we get
\begin{eqnarray}
&{\ }& \hspace{-20mm}
P(\nu_\mu \rightarrow \nu_\mu) 
\simeq 1-4|U_{\mu3}|^2(1-|U_{\mu3}|^2)
\sin^2\left(\frac{\Delta m_{31}^2 L}{4E}\right)
\nonumber\\ 
&{\ }& \hspace{1mm}
\simeq 1-\sin^22\theta_{23}
\sin^2\left(\frac{\Delta m_{31}^2 L}{4E}\right)\,.
\label{pmm}
\end{eqnarray}
Again Eq.\,(\ref{pmm}) has the same form as the oscillation
probability in the two flavor framework.
From the results of atmospheric neutrino measurements
of the Superkamiokande experiment\,\,\cite{Fukuda:1998mi},
we have
\begin{eqnarray}
&{\ }& \hspace{-70mm}
\sin^2\theta_{23} \simeq 0.5
\label{theta23}\\ 
&{\ }& \hspace{-70mm}
|\Delta m_{31}^2|\simeq 2.5\times 10^{-3}\mbox{\rm eV}^2\,.
\label{delm31}
\end{eqnarray}
Notice that the sign of $\Delta m_{31}^2$ cannot be determined
from Eq.\,(\ref{pmm}) because it is invariant under the
change of the sign of $\Delta m_{31}^2$.
The mass pattern for $\Delta m_{31}^2 > 0$ ($\Delta m_{31}^2 < 0$)
is called Normal (Inverted) Ordering, and is depicted in Fig.\,\ref{fig1}.

As for ($\theta_{12}$, $\Delta m_{21}^2$), there are two ways
to determine them.  The first one is from KamLAND\,\cite{KamLAND:2002uet},
the long baseline reactor experiment.
If we put $\theta_{13}\to 0$ again, then
in the zeroth approximation we have
$A_{\mbox{\tiny CC}}\ll|\Delta E_{21}|\simeq|\Delta m_{21}^2 /2E|\ll
|\Delta E_{31}|\simeq |\Delta m_{31}^2 /2E|$,
because the matter effect, which
appears in the form of $A_{\mbox{\tiny CC}}L\sim
L/(2000~\mbox{\rm km})$ in the
oscillation probability, can be ignored.
Hence we obtain the oscillation probability
\begin{eqnarray}
&{\ }& \hspace{-22mm}
P(\bar{\nu}_e\rightarrow\bar{\nu}_e)
\simeq 1-4|U_{e1}|^2|U_{e2}|^2\sin^2\left(\frac{\Delta m_{21}^2 L}{4E}\right)
\nonumber\\
&{\ }& \hspace{-2mm}
\simeq 1-\sin^22\theta_{12}\sin^2\left(\frac{\Delta m_{21}^2 L}{4E}\right)\,,
\label{solar3}
\end{eqnarray}
which has the same form as the oscillation
probability in the two flavor framework.
From the KamLAND result, we get
\begin{eqnarray}
&{\ }& \hspace{-62mm}
\sin^2\theta_{12} \simeq 0.3
\label{theta12kl}\\
&{\ }& \hspace{-62mm}
|\Delta m_{21}^2| \simeq 8\times 10^{-5}~\mbox{\rm eV}^2\,.
\label{delm21kl}
\end{eqnarray}
The other way to determine ($\theta_{12}$, $\Delta m_{21}^2$)
is from solar neutrino observations.
In the case of solar neutrinos, we have to take into
account the fact that the density of electrons varies
(we know now that it varies adiabatically) from the
production point of neutrinos near the center of the Sun
all the way to the Earth.
If we assume that the electron density varies adiabatically,
then the oscillation probability $P(\nu_e\rightarrow\nu_e)$
is given by
\begin{eqnarray}
&{\ }& \hspace{-0mm}
P(\nu_e\rightarrow\nu_e)=
\sum_{j,k}
\tilde{U}_{e j}(L)
\tilde{U}_{e k}^\ast(L)
\tilde{U}_{e j}^\ast(0)
\tilde{U}_{e k}(0)
\exp\left(-i\int_0^{L}dt~\Delta \tilde{E}_{jk}\right)\,,
\label{probmv}
\end{eqnarray}
where $t=0$ ($t=L$) stands for the production (detection) point,
and we discuss here detection of solar neutrinos during the day,
i.e., the density at the detection point is approximately zero.
Since the distance $L$ between the Sun and the Earth is
literally astronomically large, we have
$|\int_0^L dt~\Delta \tilde{E}_{jk}|\gg 1~(j\ne k)$,
so by averaging over rapid oscillations, we can put
$\exp(-i\int_0^L dt~\Delta \tilde{E}_{jk})\rightarrow\delta_{jk}$.
Thus we get
\begin{eqnarray}
&{\ }& \hspace{-26mm}
P(\nu_e\rightarrow\nu_e)=
\sum_{j}
\tilde{U}_{e j}(L)
\tilde{U}_{e j}^\ast(L)
\tilde{U}_{e j}^\ast(0)
\tilde{U}_{e j}(0)
\nonumber\\
&{\ }& \hspace{-5mm}
=\sum_{j}
|U_{e j}|^2
|\tilde{U}_{e j}(0)|^2
\,,
\label{probsol}
\end{eqnarray}
where the mixing matrix
$\tilde{U}(L)$ at the detection point $t=L$
was replaced by the one $U$ in vacuum
because the density there is assumed to be zero.
If take the limit $\theta_{13}\to 0$, then
Eq.\,(\ref{probsol}) reduces to the two flavor case.
In the two flavor case, the effective mixing angle
is given by
\begin{eqnarray}
&{\ }& \hspace{-5mm}
\cos2\tilde{\theta}_{12}\equiv
\frac{1}{\Delta\tilde{E}_{21}}
\left(\Delta E_{21}\cos2\theta_{12}-A_{\mbox{\tiny CC}}\right)\,,
\nonumber\\
&{\ }& \hspace{-5mm}
\Delta\tilde{E}_{21}\equiv
\left\{\left[\Delta E_{21}\cos2\theta_{12}-A_{\mbox{\tiny CC}})
\right]^2+\left(\Delta E_{21}\sin2\theta_{12}\right)^2\right\}^{1/2}\,.
\nonumber
\end{eqnarray}
Since the effective mixing matrix elements in the two flavor case
are given by $|\tilde{U}_{e 1}(0)|^2 = \cos^2\tilde{\theta}_{12}$
and $|\tilde{U}_{e 2}(0)|^2 = \sin^2\tilde{\theta}_{12}$,
we have
\begin{eqnarray}
&{\ }& \hspace{-10mm}
P(\nu_e\rightarrow\nu_e)\simeq
\frac{1}{2}
\left(1+\cos2\theta_{12}\frac{\Delta E_{21}\cos2\theta_{12}-A_{\mbox{\tiny CC}}(0)}
{\Delta\tilde{E}_{21}(0)}
\right)\,,
\label{probsol2}
\end{eqnarray}
where the argument $(0)$ stands for the quantity
evaluated at the production point of neutrinos.
From the oscillation probability (\ref{probsol2})
and the data of various experiments which measure
$P(\nu_e\rightarrow\nu_e)$ at different solar neutrino
energies, the following results are obtained\,\cite{Super-Kamiokande:2005wtt}:
\begin{eqnarray}
&{\ }& \hspace{-62mm}
\sin^2\theta_{12} \simeq 0.3
\label{theta12}\\ 
&{\ }& \hspace{-62mm}
\Delta m_{21}^2\simeq 6\times 10^{-5}\mbox{\rm eV}^2\,,
\label{delm21}
\end{eqnarray}
which are approximately consistent with the results
(\ref{theta12kl}) and (\ref{delm21kl})
from KamLAND.  Notice that the result from
solar neutrino measurements is powerful
to determine the sign of $\Delta m_{21}^2$
because of the matter effect, while
KamLAND gives little information
on the sign of $\Delta m_{21}^2$
since the oscillation probability
(\ref{solar3}) for KamLAND is basically
that in vacuum and therefore it is invariant
under the change of the sign of $\Delta m_{21}^2$.

\subsection{Parameter degeneracy}
\begin{figure}[H]
\begin{center}
\hglue -1.4cm
\includegraphics[scale=0.75]{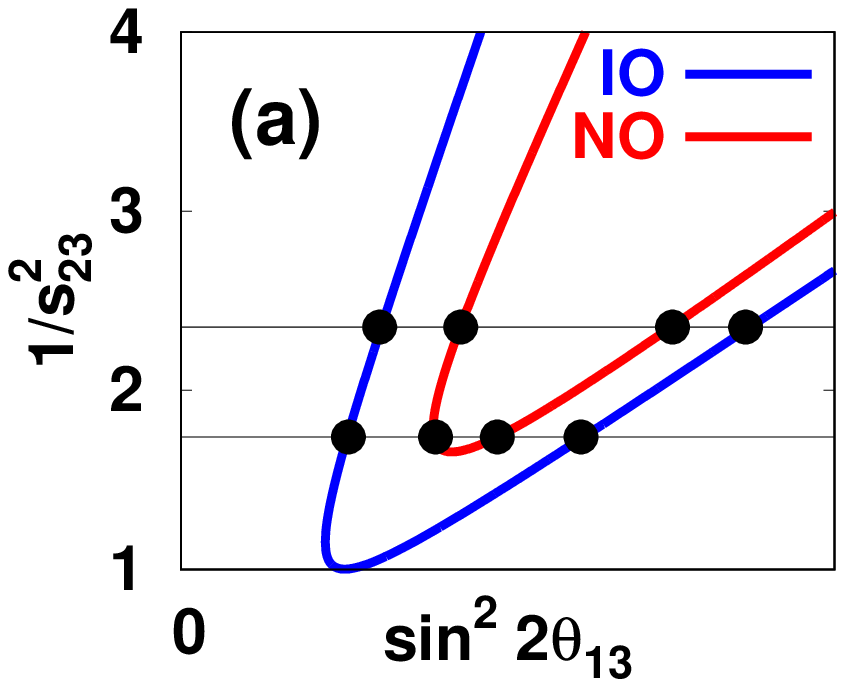}
\includegraphics[scale=0.75]{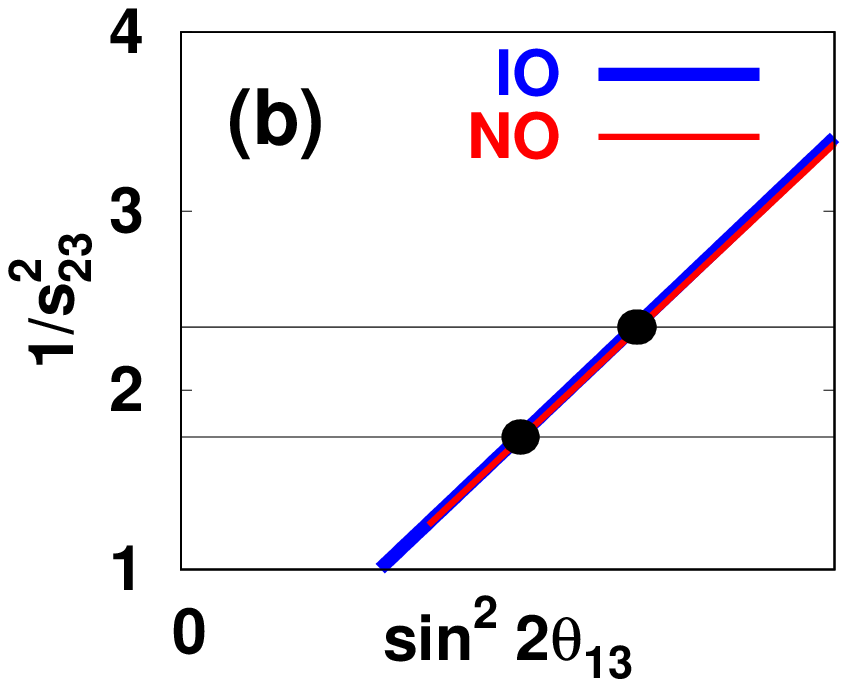}
\vglue -0.4cm
\caption{Parameter degeneracy for Normal Ordering
  and Inverted Ordering:
(a) Off the oscillation maximum ($|\Delta m^2_{31}|L/4E\ne \pi/2$),~
  (b) At the oscillation maximum ($|\Delta m^2_{31}|L/4E=\pi/2$).
  In the case of (a), there are eight solutions with a different
  value of $\delta$.
  In the case of (b), there are four solutions with two-fold degeneracy.
}
\label{fig2}
\end{center}
\end{figure}

To determine the CP phase $\delta$,
precise measurements of the oscillation
probabilities are required, since
the effect of $\delta$ appears only
in the combination of $s_{13}e^{\pm i\delta}$
as we can see from Eq.\,(\ref{mns}).
It is expected that long baseline accelerator-based
experiments are advantageous to perform
precise measurements, because one can
control the baseline length and the neutrino energy.
Conventional neutrino beams, which can be obtained
from pion decays, are $\nu_\mu$ and $\bar{\nu}_\mu$,
so $\nu_\mu \to \nu_e$ and $\bar{\nu}_\mu \to \bar{\nu}_e$
are the two major channels to determine the CP phase.
Now the question is, given the appearance probabilities
$P(\nu_\mu \to \nu_e)= \mbox{\rm constant}\equiv P$ and
$P(\bar{\nu}_\mu \to \bar{\nu}_e)= \mbox{\rm constant}\equiv \bar{P}$
in addition to the disappearance probabilities
$P(\nu_\mu \to \nu_\mu)=$ constant and
$P(\bar{\nu}_\mu \to \bar{\nu}_\nu)=$ constant,
is it possible to determine $\delta$ uniquely?
The answer to this question turns out to be negative
because there can be 8 possible values for $\delta$,
and this is called 8-fold parameter degeneracy.
In the approximation to second order in $\theta_{13}$
and $\Delta m_{21}^2$, the appearance probabilities
can be written as\,\cite{Cervera:2000kp,Barger:2001yr}
\begin{eqnarray}
&{\ }& \hspace{-5mm}
P(\nu_\mu \to \nu_e)=x^2 F^2 + 2\, \mbox{\rm sign}(\Delta m_{31}^2)\, x y F g
\cos\left[\delta+\mbox{\rm sign}(\Delta m_{31}^2)\,\Delta\right]+ y^2 g^2 = P
\label{degene1}\\
&{\ }& \hspace{-5mm}
P(\bar{\nu}_\mu \to \bar{\nu}_e) =
x^2 \bar F^2 + 2\, \mbox{\rm sign}(\Delta m_{31}^2)\, x y \bar F g
\cos\left[\delta-\mbox{\rm sign}(\Delta m_{31}^2)\,\Delta\right]
+ y^2 g^2 = \bar{P}\,,
\label{degene2}
\end{eqnarray}
where
\begin{eqnarray}
&{\ }& \hspace{-5mm}
x \equiv s_{23} \sin 2\theta_{13}
\nonumber\\
&{\ }& \hspace{-5mm}
y \equiv \alpha c_{23} \sin 2\theta_{12}
\nonumber\\
&{\ }& \hspace{-5mm}
  (F, \bar{F}) \equiv \left\{
  \begin{array}{c}
    (f, \bar{f})~~\mbox{\rm for NO}\\
    (\bar{f}, f)~~\mbox{\rm for IO}\\
  \end{array}
  \right.
\nonumber\\
&{\ }& \hspace{-5mm}
\left\{
\begin{array}{c}
  f\\
  \bar{f}
\end{array}\right\}
\equiv \frac{\sin\left(\Delta\mp A_{\mbox{\tiny CC}}L/2\right)}
{\left(1\mp A_{\mbox{\tiny CC}}L/2\Delta\right)}\,,
  \label{f}\\
&{\ }& \hspace{-5mm}
g \equiv \frac{\sin\left(A_{\mbox{\tiny CC}}L/2\right)}
{A_{\mbox{\tiny CC}}L/2\Delta}
\nonumber\\
&{\ }& \hspace{-5mm}
\Delta \equiv \frac{|\Delta m_{31}^2| L}{4E}
= 1.27
\times\frac{(|\Delta m_{31}^2|/{\rm eV^2}) (L/{\rm km})}
  {(E/{\rm GeV})}
\nonumber\\
&{\ }& \hspace{-5mm}
\alpha \equiv \left|\frac{\Delta m^2_{21}}{\Delta m^2_{31}}\right|
\nonumber
\end{eqnarray}
Defining $X\equiv\sin^22\theta_{13}$, $Y\equiv 1/s^2_{23}$,
and eliminating $\delta$, we obtain the following
expression from Eqs.\,(\ref{degene1}) and
(\ref{degene2})\,\cite{Yasuda:2004gu}:
\begin{eqnarray}
&{\ }&\hspace*{-13mm}
16CX(Y-1)
=\frac{1}{\cos^2\Delta}\left[\left(\frac{P-C}{F}
+\frac{\bar{P}-C}{\bar{F}}\right)(Y-1)-(F+\bar{F})X
+\frac{P}{F}+\frac{\bar{P}}{\bar{F}}\right]^2\nonumber\\
&{\ }&\hspace*{13mm}
+\frac{1}{\sin^2\Delta}\left[\left(\frac{P-C}{F}
-\frac{\bar{P}-C}{\bar{F}}\right)(Y-1)-(F-\bar{F})X
+\frac{P}{F}-\frac{\bar{P}}{\bar{F}}\right]^2\,,
\label{degene3}
\end{eqnarray}
where
\begin{eqnarray}
&{\ }& \hspace{-30mm}
C\equiv\left(\frac{\Delta m^2_{21}}{\Delta m^2_{31}}\right)^2
\left[\frac{\sin(A_{\mbox{\tiny CC}}L/2)}{A_{\mbox{\tiny CC}}L/2\Delta}\right]^2
\sin^22\theta_{12}\,.
\nonumber
\end{eqnarray}
Eq.\,(\ref{degene3}) gives a quadratic curve
in the $(X, Y)$-plane.
At the oscillation maximum ($|\Delta m_{31}^2| L / 4E=\pi/2$),
the numerator of the first term on the right hand side
of Eq.\,(\ref{degene3}) must vanish, and it yields
a straight line in the $(X, Y)$-plane:
\begin{eqnarray}
&{\ }& \hspace{-10mm}
\left(\frac{P-C}{F}
+\frac{\bar{P}-C}{\bar{F}}\right)(Y-1)-(F+\bar{F})X
+\frac{P}{F}+\frac{\bar{P}}{\bar{F}} = 0\,.
\label{degene4}
\end{eqnarray}
Eqs.\,(\ref{degene3}) and (\ref{degene4}) are depicted
in Fig.\,\ref{fig2} (a) and (b).
As can be seen from Fig.\,\ref{fig2},
there are eight (four) solutions in general off (at) the oscillation maximum.
This parameter degeneracy is in general eight fold\,\cite{Barger:2001yr}, since
there are two intersections between the curve \,(\ref{degene3}) and $Y$= constant
due to quadratic nature of the curve\,(\ref{degene3})
(intrinsic degeneracy\,\cite{BurguetCastell:2001ez}),
there are two curves depending on whether the mass ordering
is Normal (red in Fig.\,\ref{fig2}) or Inverted (blue)
(sign degeneracy\,\cite{Minakata:2001qm}),
and there are two possibilities for the $\theta_{23}$ octant given a value
of $\sin^22\theta_{23}$ (octant degeneracy\,\cite{Fogli:1996pv}).
The problem with this parameter degeneracy is that
the value of $\delta$ is different for a different point in Fig.\,\ref{fig2},
and we must resolve this degeneracy to determine $\delta$ precisely.
In particular, it is known\,\cite{Barger:2001yr} that the value
of $\delta$ depends strongly on the mass ordering,
so even if the long baseline experiment is done at
the oscillation maximum, resolution of sign degeneracy
is important to determine $\delta$.

\subsection{Status of measurements of the oscillation parameters}
Measurements of $\theta_{13}$ have been 
done by the appearance channel
$\nu_\mu\to\nu_e$
at the accelerator neutrino
experiments, T2K\,\cite{T2K:2011ypd}, MINOS\,\cite{Adamson:2011qu},
and by the disappearance channel
$\bar\nu_e\to\bar\nu_e$
at the reactor neutrino
experiments, Double-CHOOZ\,\cite{Abe:2011fz},
Daya Bay\,\cite{An:2012eh} and Reno\,\cite{Ahn:2012nd},
and $\theta_{13}$ has been determined precisely.
On the other hand,
the two long baseline experiments, T2K\,\cite{T2K:2021xwb}
and Nova\,\cite{NOvA:2021nfi} are performed approximately
at the oscillation maximum, so the present situation is
approximately described by (b) of Fig.\,\ref{fig2}.
\begin{table}[H]
\centering
  \begin{tabular}{|c|cc|cc|cc|}
    \hline
    \hline
    \footnotesize Ref &
    \multicolumn{2}{|c|}{\footnotesize \cite{Gonzalez-Garcia:2021dve} w SK-ATM} &
    \multicolumn{2}{|c|}{\footnotesize \cite{Capozzi:2021fjo} w SK-ATM} &
    \multicolumn{2}{|c|}{\footnotesize \cite{deSalas:2020pgw} w SK-ATM} \\
    \hline
    {\footnotesize $\begin{array}{c}\mbox{\rm Date of}\\ \mbox{\rm update}\end{array}$}&
    \multicolumn{2}{|c|}{\footnotesize 29 Nov 2021} &
    \multicolumn{2}{|c|}{\footnotesize 28 Sep 2021} &
    \multicolumn{2}{|c|}{\footnotesize 19 Jan 2021} \\
    \hline
    \footnotesize NO&
    \multicolumn{2}{|c|}{\footnotesize Best Fit Ordering} &
    \multicolumn{2}{|c|}{\footnotesize Best Fit Ordering} &
    \multicolumn{2}{|c|}{\footnotesize Best Fit Ordering} \\
    \hline
    \footnotesize Param&
    \footnotesize bfp $\pm 1\sigma$ &\footnotesize 3$\sigma$ range &
    \footnotesize bfp $\pm 1\sigma$ &\footnotesize 3$\sigma$ range &
    \footnotesize bfp $\pm 1\sigma$ &\footnotesize 3$\sigma$ range \\
    \hline
        {\footnotesize $\displaystyle
          \sin^2\theta_{12}/10^{-1}$}
& \footnotesize $3.04_{-0.12}^{+0.12}$ & \footnotesize $2.69 \to 3.43$	
& \footnotesize $3.03_{-0.13}^{+0.13}$ & \footnotesize $2.63 \to 3.45$	
& \footnotesize	$3.18_{-0.16}^{+0.16}$ & \footnotesize $ 2.71 \to 3.69$ \\
    {\footnotesize $\theta_{12}/{}^\circ$} 
& \footnotesize $33.5_{-0.8}^{+0.8}$ & \footnotesize $31.3 \to 35.9$	
& \footnotesize $33.4_{-0.8}^{+0.8}$ & \footnotesize $30.9 \to 36.0$	
& \footnotesize	$34.3_{-1.0}^{+1.0}$ & \footnotesize $ 31.4 \to 37.4$ \\
    {\footnotesize $\displaystyle
          \sin^2\theta_{23}/10^{-1}$} 
& \footnotesize $4.50_{-0.16}^{+0.19}$ & \footnotesize $4.08 \to 6.03$	
& \footnotesize $4.55_{-0.15}^{+0.18}$ & \footnotesize $4.16 \to 5.99$	
& \footnotesize	$5.74_{-0.14}^{+0.14}$ & \footnotesize $ 4.34 \to 6.10$ \\
    {\footnotesize $\theta_{23}/{}^\circ$} 
& \footnotesize $42.1_{-0.9}^{+1.1}$ & \footnotesize $39.7 \to 50.9$	
& \footnotesize $42.4_{-0.9}^{+1.0}$   & \footnotesize $40.2 \to 50.7$	
& \footnotesize	$49.3_{-0.8}^{+0.8}$ & \footnotesize $ 41.2 \to 51.3$ \\
    {\footnotesize $\displaystyle
          \sin^2\theta_{13}/10^{-2}$} 
& \footnotesize $2.25_{-0.06}^{+0.06}$ & \footnotesize $2.06 \to 2.44$	
& \footnotesize $2.23_{-0.06}^{+0.07}$ & \footnotesize $2.04 \to 2.44$	
& \footnotesize	$2.20^{+0.07}_{-0.06}$ & \footnotesize $ 2.00 \to 2.41$ \\
    {\footnotesize $\theta_{13}/{}^\circ$}
& \footnotesize $8.62_{-0.12}^{+0.12}$ & \footnotesize $8.25 \to 8.98$	
& \footnotesize $8.59_{-0.12}^{+0.13}$ & \footnotesize $8.21 \to 8.99$	
& \footnotesize	$8.53^{+0.13}_{-0.12}$ & \footnotesize $ 8.13 \to 8.92$ \\
    {\footnotesize $\delta/{}^\circ$}
& \footnotesize $230_{-25}^{+36}$ & \footnotesize $144 \to 350$		
& \footnotesize $274_{-27}^{+25}$  & \footnotesize $139 \to 355$	
& \footnotesize	$194^{+24}_{-22}$ & \footnotesize $ 128 \to 359$ \\
    {\footnotesize $\displaystyle{\Delta m^2_{21}}/{10^{-5}~\mbox{\rm eV}^2}$}
& \footnotesize $7.42_{-0.20}^{+0.21}$ & \footnotesize $6.82 \to 8.04$	
& \footnotesize $7.36_{-0.15}^{+0.16}$ & \footnotesize $6.93 \to 7.93$	
& \footnotesize	$7.50^{+0.22}_{-0.20}$ & \footnotesize $ 6.94 \to 8.14$ \\
   {\footnotesize $\displaystyle{\Delta m^2_{\mbox{\tiny\rm atm}}}/{10^{-3}~\mbox{\rm eV}^2}$}
& \footnotesize $2.51_{-0.03}^{+0.03}$ & \footnotesize $2.43 \to 2.59$	
& \footnotesize $2.49_{-0.03}^{+0.02}$ & \footnotesize $2.40 \to 2.57$	
& \footnotesize	$2.55^{+0.02}_{-0.03}$ & \footnotesize $ 2.47 \to 2.63$ \\
    \hline
    \footnotesize IO &
    \multicolumn{2}{|c|}{\footnotesize $\Delta\chi^2$=7.0} &
    \multicolumn{2}{|c|}{\footnotesize $\Delta\chi^2$=6.5} &
    \multicolumn{2}{|c|}{\footnotesize $\Delta\chi^2$=6.4} \\
    \hline
        {\footnotesize $\displaystyle
          \sin^2\theta_{12}/10^{-1}$}
& \footnotesize $3.04_{-0.12}^{+0.13}$ & \footnotesize $2.69 \to 3.43$	
& \footnotesize $3.03_{-0.13}^{+0.13}$ & \footnotesize $2.63 \to 3.45$	
& \footnotesize	$3.18_{-0.16}^{+0.16}$ & \footnotesize $ 2.71 \to 3.69$ \\
   {\footnotesize $\theta_{12}/{}^\circ$}
& \footnotesize $33.5_{-0.8}^{+0.8}$ & \footnotesize $31.3 \to 35.9$	
& \footnotesize $33.4_{-0.8}^{+0.8}$ & \footnotesize $30.9 \to 36.0$	
& \footnotesize	$34.3_{-1.0}^{+1.0}$ & \footnotesize $ 31.4 \to 37.4$ \\
    {\footnotesize $\displaystyle
          \sin^2\theta_{23}/10^{-1}$} 
& \footnotesize $5.70_{-0.22}^{+0.16}$ & \footnotesize $4.10 \to 6.13$	
& \footnotesize $5.69_{-0.21}^{+0.13}$ & \footnotesize $4.17 \to 6.06$	
& \footnotesize	$5.78^{+0.10}_{-0.17}$ & \footnotesize $ 4.33 \to 6.08$ \\
   {\footnotesize $\theta_{23}/{}^\circ$}
& \footnotesize $49.0_{-1.3}^{+0.9}$ & \footnotesize $39.8 \to 51.6$	
& \footnotesize $49.0_{-1.2}^{+0.8}$   & \footnotesize $40.22 \to 51.12$	
& \footnotesize	$49.5^{+0.6}_{-1.0}$ & \footnotesize $ 41.2 \to 51.3$ \\
    {\footnotesize $\displaystyle
          \sin^2\theta_{13}/10^{-2}$} 
& \footnotesize $2.24_{-0.06}^{+0.07}$ & \footnotesize $2.06 \to 2.46$	
& \footnotesize $2.23_{-0.06}^{+0.06}$ & \footnotesize $2.03 \to 2.45$	
& \footnotesize	$2.23^{+0.06}_{-0.07}$ & \footnotesize $ 2.02 \to 2.42$ \\
   {\footnotesize $\theta_{13}/{}^\circ$}
& \footnotesize $8.61_{-0.12}^{+0.14}$ & \footnotesize $8.24 \to 9.02$	
& \footnotesize $8.59_{-0.12}^{+0.12}$ & \footnotesize $8.19 \to 9.01$	
& \footnotesize	$8.58^{+0.12}_{-0.14}$ & \footnotesize $ 8.17 \to 8.96$ \\
   {\footnotesize $\delta/{}^\circ$}
& \footnotesize $278_{-30}^{+22}$ & \footnotesize $194 \to 345$		
& \footnotesize $274_{-27}^{+25}$  & \footnotesize $193 \to 342$	
& \footnotesize	$284^{+26}_{-28}$ & \footnotesize $ 200 \to 353$ \\
   {\footnotesize $\displaystyle{\Delta m^2_{21}}/{10^{-5}~\mbox{\rm eV}^2}$}
& \footnotesize $7.42_{-0.20}^{+0.21}$ & \footnotesize $6.82 \to 8.04$	
& \footnotesize $7.36_{-0.15}^{+0.16}$ & \footnotesize $6.93 \to 7.93$	
& \footnotesize	$7.50^{+0.22}_{-0.20}$ & \footnotesize $ 6.94 \to 8.14$ \\
   {\footnotesize $\displaystyle{\Delta m^2_{\mbox{\tiny\rm atm}}}/{10^{-3}~\mbox{\rm eV}^2}$}
& \footnotesize $2.49_{-0.03}^{+0.03}$ & \footnotesize $2.41 \to 2.57$	
& \footnotesize $2.46_{-0.03}^{+0.03}$ & \footnotesize $2.38 \to 2.54$	
& \footnotesize	$2.45^{+0.02}_{-0.03}$ & \footnotesize $ 2.37 \to 2.53$ \\
    \hline
{\footnotesize Def of $\Delta m^2_{\mbox{\tiny\rm atm}}$}&
\multicolumn{2}{|c|}{\footnotesize $\max\left(|\Delta m^2_{31}|,|\Delta m^2_{32}|\right)$} &
\multicolumn{2}{|c|}{\footnotesize $|m^2_3-(m^2_1+m^2_2)/2|$} &
\multicolumn{2}{|c|}{\footnotesize $|\Delta m^2_{31}|$} \\
    \hline
    \hline
  \end{tabular}
\vspace{5mm}
  \caption{The updated results of global fit by the three groups.
    The definition of the mass squared difference $\Delta m^2_{\mbox{\scriptsize\rm atm}}$
    of atmospheric neutrino oscillation is different for
    different groups.  Although 
    two different results, with and without the Superkamiokande
    atmospheric neutrino data, are presented
    in Ref.\,\cite{Gonzalez-Garcia:2021dve}, only those with
    the atmospheric neutrino data are quoted.
    Normal ordering gives the best fit, but the significance of
    NO over IO is not strong enough to exclude IO yet.}
\label{tab1}
\end{table}
Table \ref{tab1} gives the results by the three
groups\,\cite{deSalas:2020pgw,Capozzi:2021fjo,Gonzalez-Garcia:2021dve}
for the values of the mixing angles and the mass
squared differences from global analysis of
solar neutrinos
(Chlorine\,\cite{Cleveland:1998nv},
Gallex/GNO\,\cite{Kaether:2010ag}, SAGE\,\cite{SAGE:2009eeu},
and Super-Kamiokande\,\cite{Super-Kamiokande:2005wtt,Super-Kamiokande:2008ecj,Super-Kamiokande:2010tar,SK:nu2020}, SNO\,\cite{SNO:2011hxd}, and
Borexino\,\cite{Bellini:2011rx,Borexino:2008fkj,BOREXINO:2014pcl}.)
KamLAND\,\cite{KamLAND:2013rgu},
atmospheric neutrinos
(Super-Kamiokande\,\cite{Super-Kamiokande:2017yvm} and
IceCube/DeepCore\,\cite{IceCube:2014flw,deepcore:2016})
medium baseline reactor neutrinos (Double Chooz\,\cite{DoubleC:nu2020},
Daya Bay\,\cite{DayaBay:2018yms},
and RENO\,\cite{RENO:nu2020}),
disappearance channel at accelerator-based long baseline experiments
(MINOS\,\cite{MINOS:2013utc}, T2K\,\cite{T2K:nu2020}, and NOvA\,\cite{NOvA:nu2020}),
and appearance channel at accelerator-based long baseline experiments
(MINOS\,\cite{MINOS:2013xrl}, T2K\,\cite{T2K:nu2020}, and NOvA\,\cite{NOvA:nu2020}).
As can be seen from Eq.\,(\ref{f}), the matter effect
appear in the form $A_{\mbox{\tiny CC}}L/2\sim L/4000$ km.
Since the baseline lengths of T2K and Nova
are not long enough to satisfy $A_{\mbox{\tiny CC}}L/2\sim O(1)$,
the results by these long baseline experiments
are not conclusive enough to determine the mass pattern
as of 2021.
Octant degeneracy is also unresolved.

\subsection{Prospect of future experiments}

\begin{figure}[H]
\begin{center}
\begin{tabular}{lr}
\vspace{0.2 in}
\hspace{-0.35 in}
\hglue -1.4cm
\includegraphics[width=0.5\textwidth]{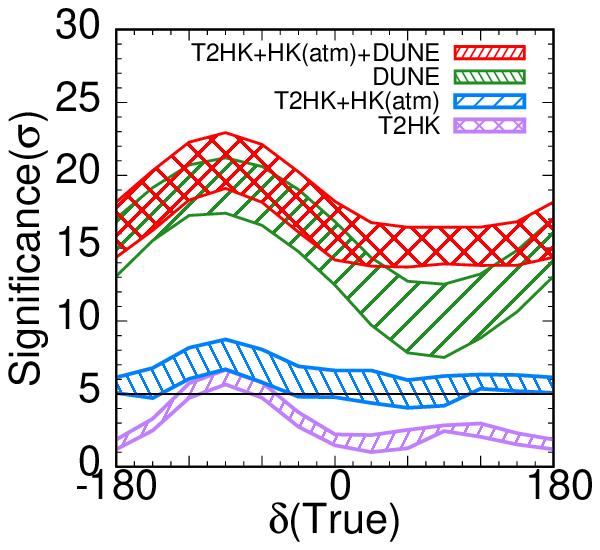}
\hspace{-1.0 in}
\includegraphics[width=0.5\textwidth]{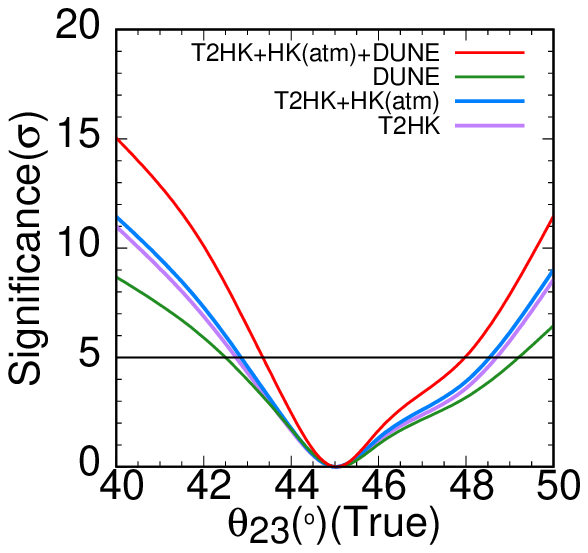}
\hspace{-1.0 in}
\includegraphics[width=0.5\textwidth]{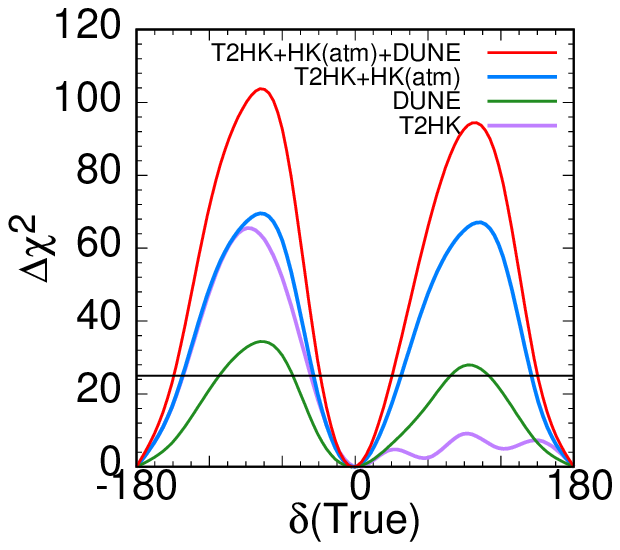} 
\end{tabular}
\end{center}
\caption{Sensitivity of the future experiments and their combinations
to the unknown quantities in the case of normal
ordering\,\cite{Fukasawa:2016yue}.  Left panel: Significance of
mass ordering as a function of the true value of the CP phase
$\delta$.  Middle panel: Significance of octant as a function of the
true value of $\theta_{23}$.  Right panel: $\Delta\chi^2=\{\mbox{\rm
significance of CP violation}(\sigma)\}^2$ as a function of
the true value of the CP phase $\delta$.  The horizontal
straight line in each panel stands for $5\sigma$ which corresponds
to discovery.  }
\label{fig3}
\end{figure}
The quantities which are not determined as of 2021 are
the pattern of mass ordering (the two mass
patterns in Fig.\ref{fig1} are allowed at present),
the octant of $\theta_{23}$ (in other words
the sign of $(\theta_{23}-45^\circ)$) and the CP phase $\delta$.
Determination of
the mass ordering,
the octant of $\theta_{23}$ and the CP phase $\delta$
is expected to be carried out
in the future long baseline experiments,
T2HK\,\cite{Hyper-KamiokandeProto-:2015xww}
and DUNE\,\cite{Acciarri:2015uup}.
The figure \ref{fig3} shows the
sensitivity of these experiments together with
atmospheric neutrino measurements at the
Hyperkamiokande (HK) experiment\,\cite{Abe:2011ts}
to the unknown quantities.
The DUNE experiment has a baseline length $L$=1300 km,
and it has the highest sensitivity to the mass ordering
and the significance to reject wrong mass ordering is
at least 7$\sigma$ for any value of $\delta$.
T2HK has a short baseline length $L$=295 km,
and it has therefore poor sensitivity to the mass
ordering, but if it is combined with the HK atmospheric
neutrino data, then the significance to 
reject wrong mass ordering is at least 4$\sigma$.
Octant degeneracy is expected to be solved
if $|\theta_{23}-45^\circ|>3^\circ$ if we combine
T2HK, DUNE and HK atmospheric neutrinos.
As for the CP phase, unless $\delta$ is close
to 0 or 180$^\circ$, DUNE, the combination of T2HK and atmospheric
neutrino observation at HK and the combination of all these can
exclude the hypothesis $\delta=0$ or $\delta=$180$^\circ$.

\section{New physics beyond the Standard Model with massive three neutrinos\label{np}}

As of 2021, there are several anomalies which cannot be
accounted for by the standard three-flavor framework.
One is a class of anomalies which may be explained
if we assume the existence of
light sterile neutrinos\,\cite{Abazajian:2012ys} whose
mass is of order 1 eV.
They are the results of LSND\,\cite{Aguilar:2001ty},
MiniBooNE\,\cite{MiniBooNE:2020pnu},
reactor antineutrino anomaly\,\cite{Mueller:2011nm,Huber:2011wv}
and gallium anomaly\,\cite{Giunti:2007xv}.
Also the recent IceCube data\,\cite{IceCube:2020phf}
suggests a mild preference for light sterile neutrinos
in the disappearance channel $\nu_\mu\to\nu_\mu$
+ $\bar\nu_\mu\to\bar\nu_\mu$.
The other anomaly is the discrepancy\,\cite{Super-Kamiokande:2016yck} between
the mass squared differences of the solar and
KamLAND experiments.
This may be explained\,\cite{Gonzalez-Garcia:2013usa,Maltoni:2015kca}
if we assume either flavor-dependent nonstandard
interactions during the propagation of
neutrinos\,\cite{Wolfenstein:1977ue,Guzzo:1991hi,Roulet:1991sm},
or light sterile neutrinos whose mass squared difference
is of order $10^{-5}$ eV$^2$.
These anomalies have attracted a lot of attention
because they may provide us a key to physics
beyond the Standard Model.

\subsection{Light sterile neutrinos}

\begin{figure}[H]
\begin{center}
\hglue -1.4cm
\includegraphics[scale=0.35]{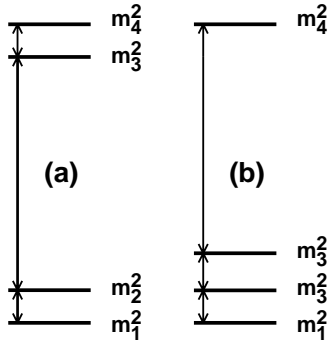}
\vglue -0.4cm
\caption{Two schemes of four neutrino
mixing.
(a): (2+2)-scheme,~
(b): (3+1)-scheme.
}
\label{fig4}
\end{center}
\end{figure}

If we assume four flavor and mass eigenstates,
then the fourth flavor eigenstate must be
sterile neutrino state $\nu_s$, which is
singlet with respect to the gauge group of the Standard Model
because the number of weakly interacting light
neutrinos is three from the LEP data\,\cite{ParticleDataGroup:2020ssz}.
So the equation of motion is described by
the following $4 \times 4$ Hamiltonian:
\begin{eqnarray}
&{\ }& \hspace{-3mm}
  \displaystyle i {d \over dt} \left( \begin{array}{c}
  \nu_e(t) \\
  \nu_\mu(t)\\
  \nu_\tau(t)\\
  \nu_s(t)
\end{array} \right)
=M~\left( \begin{array}{c} \nu_e(t) \\
\nu_\mu(t)\\\nu_\tau(t)\\
\nu_s(t)
\end{array} \right)\qquad\qquad\qquad\qquad
\label{sch2}\\
&{\ }& \hspace{-3mm}
M\equiv[U\,\mbox{\rm diag}(E_1,E_2,E_3,E_4)\,U^{-1}
  +\mbox{\rm diag}(A_{\mbox{\tiny CC}},0,0,-A_{\mbox{\tiny NC}})+A_{\mbox{\tiny NC}}\,{\bf 1}]
\nonumber\\
&{\ }& \hspace{3mm}
=[U\,\mbox{\rm diag}(0,\Delta E_{21},\Delta E_{31},\Delta E_{41})\,U^{-1}
  +\mbox{\rm diag}(A_{\mbox{\tiny CC}},0,0,-A_{\mbox{\tiny NC}})
  +(E_1+A_{\mbox{\tiny NC}})\,{\bf 1}]\,,
\label{m4}
\end{eqnarray}
The $4\times 4$ mixing matrix $U$ has 6 mixing angles and
3 CP phases, and one of the parametrization for $U$ is given by
\begin{eqnarray}
    U =
    R_{34}(\theta_{34} ,\, \delta_{34}) \; R_{24}(\theta_{24} ,\, 0) \;
    R_{14}(\theta_{14} ,\, \delta_{14}) \; R_{23}(\theta_{23} ,\, 0) \;
     R_{13}(\theta_{13} ,\, \delta_{13}) \; 
    R_{12}(\theta_{12} ,\, 0) \,,
    \label{eq:3+1param2}
\end{eqnarray}
where $R_{ij}(\theta_{ij},\ \delta_l)$ are the complex
rotation matrices in the $ij$-plane defined as:
\begin{eqnarray}
  &{\ }&\hspace*{-65mm}
  [R_{ij}(\theta_{ij},\ \delta_{ij})]_{pq} = \left\{
  \begin{array}{ll}
    \cos \theta_{ij} & p=q=i,j \\
1 & p=q \not= i,j \\
\sin \theta_{ij} \ e^{-i\delta_{ij}} &   p=i;q=j \\
-\sin \theta_{ij} \ e^{i\delta_{ij}} & p=j;q=i \\
0 & \mbox{\rm otherwise.}
  \end{array} \right.
  \nonumber
\end{eqnarray}
The angles $\theta_{14}$, $\theta_{24}$,
$\theta_{34}$ stand for the mixing
of oscillations at short baseline experiments of reactor neutrino
($\bar{\nu}_e\to\bar{\nu}_e$) and radioactive sources
($\nu_e\to\nu_e$), that of
oscillations at short baseline accelerator neutrino experiments
($\nu_\mu\to\nu_\mu$ and $\bar{\nu}_\mu\to\bar{\nu}_\mu$),
and the ratio between $\nu_\mu\to\nu_\tau$
and $\nu_\mu\to\nu_s$ at short baseline accelerator neutrinos
experiments, respectively.
In the limit $\theta_{j4}\to 0~(j=1,2,3)$,
the mixing is reduced to the PMNS paradigm
with $\delta = \delta_{13}$.
In the case of four neutrino mixing scenarios,
there are two schemes, the (2+2)-scheme
(Fig.\ref{fig4} (a)) and the
(3+1)-scheme (Fig.\ref{fig4} (b)).
The (2+2)-scheme is excluded by
combining the constraints from
the solar and atmospheric neutrinos\,\cite{Maltoni:2004ei}.
On the other hand, in the (3+1)-scheme,
three massive states must be in the lower side
as in Fig.\ref{fig2} (b) because
the one with three massive states
in the upper side is excluded by
the cosmological bound\,\cite{ParticleDataGroup:2020ssz}
$\sum_j m_j \lesssim 0.5$ eV (95\% C.L.).

If the neutrino energy $E$ and the baseline length $L$
satisfy $|\Delta m^2_{41}L/E|\sim O(1)$
for the value $\Delta m^2_{41}\gtrsim$ 0.1 eV$^2$,
then the contributions from the smaller mass
squared differences $\Delta m^2_{31}$ and $\Delta m^2_{21}$
become negligible, and the oscillation probability in vacuum
can be expressed as
\begin{eqnarray}
P(\nu_{\alpha} \to \nu_{\beta})
\simeq
P(\bar\nu_{\alpha} \to \bar\nu_{\beta})
\simeq 
\left|\delta_{\alpha \beta} - 
\sin^2{2\theta_{\alpha \beta}} 
\,\sin^2\left( \frac{\Delta{m}^2_{41} L}{4E} \right)\right|\,,
\nonumber
\end{eqnarray}
where 
\begin{eqnarray}
\sin^2{2\theta_{\alpha \beta}}
\equiv
4|U_{\alpha 4}|^2
\left| \delta_{\alpha \beta} - |U_{\beta 4}|^2 \right|
\qquad
(\alpha, \beta = e, \mu, \tau, s) \,,
\nonumber
\end{eqnarray}
so the formula for the oscillation probability reduces
to that in the two flavor framework.
In the following subsections, we will discuss the
three channels
($\alpha$, $\beta$) = ($e$, $e$), ($\mu$, $e$) and ($\mu$, $\mu$).
In the parametrization of Eq.\,(\ref{eq:3+1param2}),
we have the following mixing angle:
\begin{eqnarray}
  &{\ }&\hspace*{-20mm}
\sin^2{2\theta_{ee}}
=4|U_{e4}|^2\left(1-|U_{e4}|^2\right)
=\sin^22\theta_{14}
\label{ee4}\\
  &{\ }&\hspace*{-20mm}
\sin^2{2\theta_{\mu\mu}}
=4|U_{\mu 4}|^2\left(1-|U_{\mu 4}|^2\right)
=4c^2_{14}s^2_{24}\left(1-c^2_{14}s^2_{24}\right)
\label{mm4}\\
  &{\ }&\hspace*{-20mm}
\sin^2{2\theta_{\mu e}}
=4|U_{\mu 4}|^2|U_{e4}|^2
=s^2_{24}\sin^22\theta_{14}
\label{me4}
\end{eqnarray}

\subsubsection{$\nu_e\to\nu_e$ channel}
{\ }\\
\vglue -0.4cm
To probe neutrino oscillations $\nu_e\to\nu_e$
or $\bar\nu_e\to\bar\nu_e$ for the region
of 0.1 eV$^2\lesssim\Delta m^2\lesssim$ 10 eV$^2$,
short baseline experiments of electron antineutrinos from reactors
or electron neutrinos from radioactive sources have been performed.
In 2011 the
flux of the reactor neutrino was recalculated in
Ref.\,\cite{Mueller:2011nm,Huber:2011wv} and it was claimed
that the normalization is shifted by about +3\% on average.
If their claim on the reactor neutrino flux is correct,
then neutrino oscillation with $\Delta m^2\gtrsim$ 1 eV$^2$
may be concluded from a re-analysis of 19 reactor neutrino
results at short baselines\,\cite{Mention:2011rk}.
This is called reactor antineutrino anomaly.
The result in 2019 by Daya Bay\,\cite{Adey:2019ywk}
disfavors the new flux of reactor
antineutrinos\,\cite{Mention:2011rk,Huber:2011wv} from the shape
analysis of the energy spectrum.  However, as far as the overall
normalization of reactor antineutrinos is concerned, there is
uncertainty which is as large as that of the difference between the
old and new flux\,\cite{Hayes:2013wra} and it is not clear whether
the reactor antineutrino anomaly is disfavored from the result of
Ref.\,\cite{Adey:2019ywk}.
Among the recent short baseline reactor neutrino experiments,
NEOS\,\cite{NEOS:2016wee},
DANSS\,\cite{DANSS:2018fnn},
STEREO\,\cite{STEREO:2019ztb},
PROSPECT\,\cite{PROSPECT:2020sxr} and
Neutrino-4\,\cite{Serebrov:2020kmd},
the only experiment which had an affirmative result
is Neutrino-4.

On the other hand, it was pointed out in Ref.\,\cite{Giunti:2007xv} that
the measured and predicted $^{71}$Ge production rates
differ in the gallium radioactive source experiments GALLEX and SAGE,
and this is called gallium anomaly.  The recent affirmative result of the BEST
experiment\,\cite{Barinov:2021asz} is consistent with the region
in ($\sin^22\theta_{14}$, $\Delta m^2_{41}$) which was suggested by
Ref.\,\cite{Giunti:2007xv}, although most of the BEST region is
disfavored by the reactor experiments except Neutrino-4.

\begin{figure}
\hglue 1.7cm
\hglue -2.4cm
\includegraphics[scale=1.0]{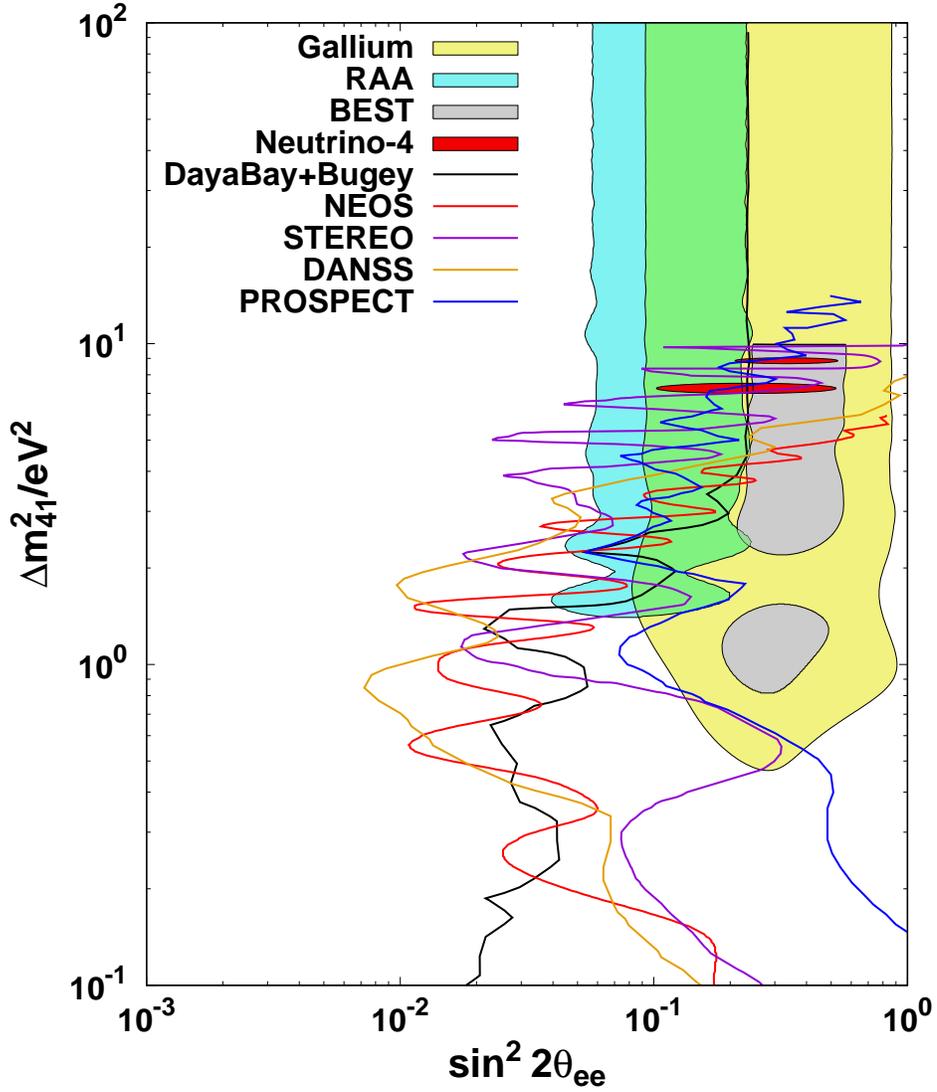}
\vglue -0.4cm
\caption{The excluded and allowed region for $\nu_e\to\nu_e$
  or $\bar{\nu}_e\to\bar{\nu}_e$ channel.
  The curves of
  DayaBay+Bugey (90\%CL)\,\cite{DayaBay:2016lkk},
  NEOS (90\%CL)\,\cite{NEOS:2016wee},
  DANSS (90\%CL)\,\cite{DANSS:2018fnn},
  STEREO (95\%CL)\,\cite{STEREO:2019ztb} and
  PROSPECT (95\%CL)\,\cite{PROSPECT:2020sxr}
  stand for negative results.
  Those of
  gallium anomaly (95\%CL)\,\cite{Giunti:2007xv},
  reactor antineutrino anomaly (RAA) (95\%CL)\,\cite{Mention:2011rk,Huber:2011wv},
  Neutrino-4 (90\%CL)\,\cite{Serebrov:2020kmd} and
  BEST (95\%CL)\,\cite{Barinov:2021asz}
  are affirmative ones.
}
\label{fig5}
\end{figure}

Fig.\ref{fig5} shows the results of reactor and radioactive
source neutrino experiments.
There are criticisms on significance
of the affirmative results and the readers are
referred to Ref.\,\cite{Athar:2021xsd} for references.
To observe the oscillation pattern
at short baseline experiments with low energy,
it is necessary to have a detector with very good energy
resolution and to have relatively small size of
a reactor core.  This is the reason why short baseline
experiments become difficult for
$\Delta m^2_{41}\gtrsim$ several eV$^2$.
In Ref.\,\cite{Wang:2021gox} it was suggested that
we may be able to improve sensitivity to $\sin^2{2\theta_{ee}}$
by observing high energy atmospheric neutrino shower events
at a neutrino telescope whose size is at least ten times as
large as the IceCube facility.
The reason why observation of high energy neutrinos
with a very long baseline length
is advantageous is because there can be
enhancement of the effective mixing angle in matter
and therefore the sensitivity can be improved.

\subsubsection{$\nu_\mu\to\nu_\mu$ channel}
{\ }\\
\vglue -0.4cm
Neutrino oscillations $\nu_\mu\to\nu_\mu$
or $\bar\nu_\mu\to\bar\nu_\mu$ for the region
of 0.1 eV$^2\lesssim\Delta m^2\lesssim$ 100 eV$^2$
have been searched by experiments with accelerator
(CDHSW\,\cite{Dydak:1983zq},
CCFR\,\cite{Stockdale:1984cg},
MiniBooNE/SciBooNE\,\cite{SciBooNE:2011qyf} and
MINOS/MINOS+\,\cite{MINOS:2020iqj})
as well as atmospheric
(Superkamiokande\,\cite{Super-Kamiokande:2014ndf}
and IceCube\,\cite{IceCube:2020phf})
neutrinos.
Most of these experiments had negative results,
and the only experiment which obtained
an affirmative result is IceCube\,\cite{IceCube:2020phf},
although its significance is weak, i.e.,
no sterile neutrino hypothesis is disfavored only
by $\Delta \chi^2=4.94$ for
two degrees of freedom, naively corresponding to 1.7$\sigma$
in the Gaussian distribution.

\begin{figure}
\hglue 0.9cm
\hglue -2.4cm
\includegraphics[scale=1.2]{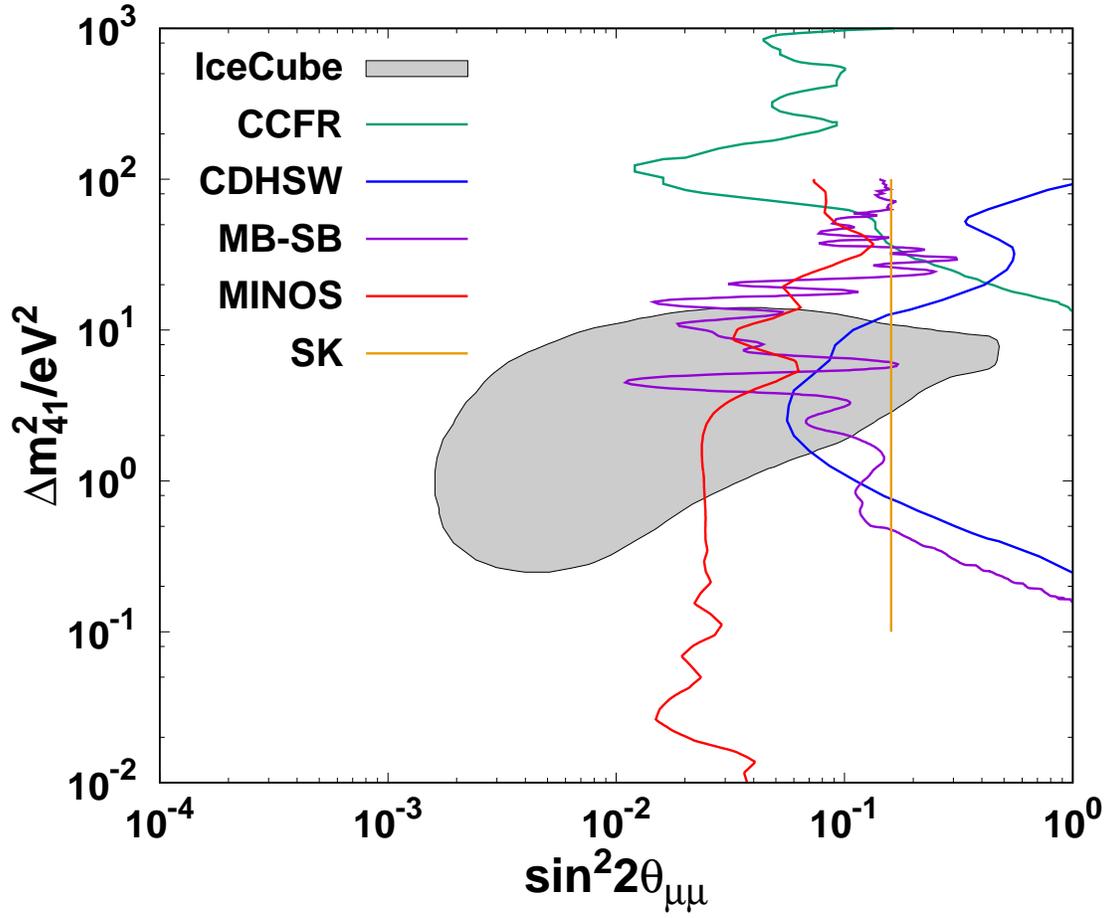}
\vglue -0.4cm
\caption{The excluded and allowed region for $\nu_\mu\to\nu_\mu$
  or $\bar{\nu}_\mu\to\bar{\nu}_\mu$ channel.
  The curves of
  CDHSW (90\%CL)\,\cite{Dydak:1983zq},
  CCFR (90\%CL)\,\cite{Stockdale:1984cg},
  MiniBooNE/SciBooNE (90\%CL)\,\cite{SciBooNE:2011qyf},
  Superkamiokande (90\%CL)\,\cite{Super-Kamiokande:2014ndf} and
  MINOS/MINOS+ (90\%CL)\,\cite{MINOS:2020iqj}
  stand for negative results.
  That of
  IceCube (90\%CL)\,\cite{IceCube:2020phf}
  is affirmative one.
}
\label{fig6}
\end{figure}

Fig.\ref{fig6} shows the results of experiments
for disappearance channel of $\nu_\mu$ or $\bar\nu_\mu$.

\subsubsection{$\nu_\mu\to\nu_e$ channel}
\begin{figure}
\begin{center}
\hglue -2.4cm
\includegraphics[scale=1.2]{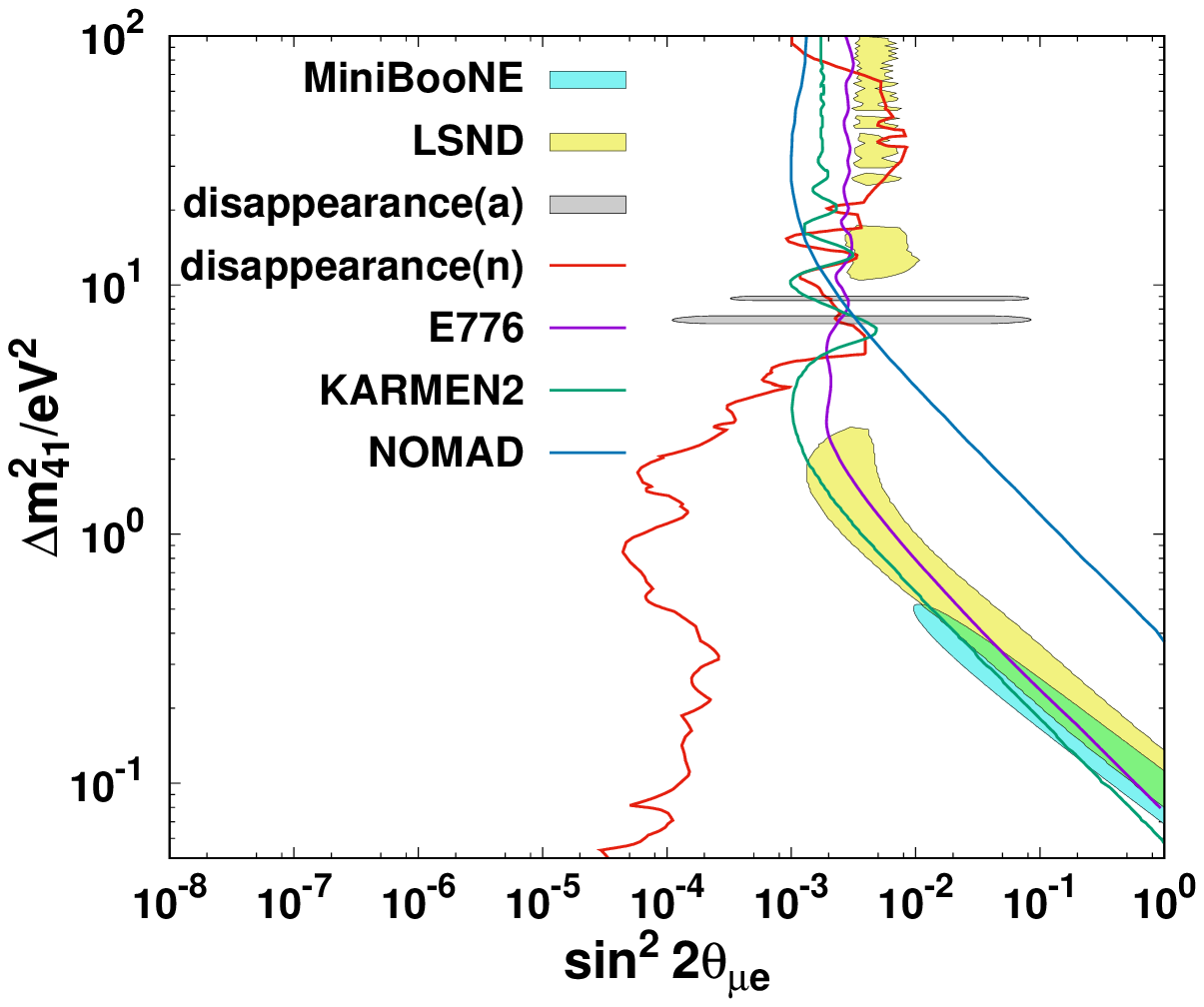}
\vglue -0.4cm
\caption{The excluded and allowed region for $\nu_\mu\to\nu_e$ or
  $\bar{\nu}_\mu\to\bar{\nu}_e$ channel.  The curves of 
  E776 (90\%CL)\,\cite{Borodovsky:1992pn},
  KARMEN2 (90\%CL)\,\cite{KARMEN:2002zcm} and
  NOMAD (90\%CL)\,\cite{NOMAD:2003mqg}
  stand for negative results.  Those of 
  LSND (90\%CL)\,\cite{Aguilar:2001ty} and
  MiniBooNE (90\%CL)\,\cite{MiniBooNE:2020pnu}
  are affirmative ones.
  ``Disappearance(n)'' stands for the constraint at 90\%CL from the
  negative results of disappearance channels $\nu_e\to\nu_e$
  ($\bar{\nu}_e\to\bar{\nu}_e$) and $\nu_\mu\to\nu_\mu$
  ($\bar{\nu}_\mu\to\bar{\nu}_\mu$) as is given by Eq.\,(\ref{me4v2}).
  ``Disappearance(a)'' stands for the possible region at 90\%CL which
  can be inferred if we take seriously the two affirmative results of
  disappearance channels, i.e., Neutrino-4
  ($\bar{\nu}_e\to\bar{\nu}_e$) and IceCube ($\nu_\mu\to\nu_\mu$
  and/or $\bar{\nu}_\mu\to\bar{\nu}_\mu$).  }
\label{fig7}
\end{center}
\end{figure}

{\ }\\
\vglue -0.4cm
Several accelerator-based neutrino experiments
have been performed in the past to search for oscillations
$\nu_\mu\to\nu_e$ or $\bar\nu_\mu\to\bar\nu_e$.
The only two experiments which reported an affirmative result
are LSND\,\cite{Aguilar:2001ty} and MiniBooNE\,\cite{MiniBooNE:2020pnu}.
Since the early stage of sterile neutrino oscillation study,
it has been known\,\cite{Okada:1996kw,Bilenky:1996rw} that the LSND region in the
($\Delta m^2$, $\sin^22\theta$)-plane has tension with
other negative results in the disappearance channels
$\nu_e\to\nu_e$ and $\nu_\mu\to\nu_\mu$.
By defining $\sin^22\theta_{\alpha\beta}~(\alpha,\beta=\mu, e)$
as a value of the horizontal coordinate
as a function of $\Delta m^2_{41}$ in Figs.\ref{fig5} and \ref{fig5},
from Eqs.\,(\ref{ee4}) and (\ref{mm4}),
negative results in the two channels
$\nu_e\to\nu_e$ and $\nu_\mu\to\nu_\mu$ at a given value of $\Delta m^2_{41}$ indicate
\begin{eqnarray}
  &{\ }&\hspace*{-40mm}
  4|U_{e4}|^2\left(1-|U_{e4}|^2\right) < \sin^22\theta_{ee}
\label{ee4v2}
  \nonumber\\
  &{\ }&\hspace*{-40mm}
  4|U_{\mu 4}|^2\left(1-|U_{\mu 4}|^2\right) < \sin^22\theta_{\mu\mu}\,.
\label{mm4v2}
\end{eqnarray}
On the other hand, from Eq.\,(\ref{me4}),
an affirmative result in the channel $\nu_\mu\to\nu_e$
leads to
\begin{eqnarray}
  &{\ }&\hspace*{-40mm}
  4|U_{\mu 4}|^2|U_{e4}|^2 = \sin^22\theta_{\mu e}\,.
\label{me4v2}
\end{eqnarray}
Assuming $|U_{e4}|^2\ll 1$, $|U_{\mu 4}|^2\ll 1$,
we have the following inequality at a given value of $\Delta m^2_{41}$:
\begin{eqnarray}
  &{\ }&\hspace*{-30mm}
  \sin^22\theta_{\mu e} < \frac{1}{4}\sin^22\theta_{ee}
  \sin^22\theta_{\mu\mu}\,.
\label{me4v3}
\end{eqnarray}
$\sin^22\theta_{ee}$ ($\sin^22\theta_{\mu\mu}$)
can be regarded as the
strongest bound from the negative results of
the disappearance channel $\nu_e\to\nu_e$
($\nu_\mu\to\nu_\mu$).
The value of $(1/4)\sin^22\theta_{ee}\sin^22\theta_{\mu\mu}$
at 90\%CL
is plotted as a function of $\Delta m^2_{41}$
in Fig.\,\ref{fig7} (indicated as ``disappearance(n)'')
together with the allowed regions suggested by
LSND and MiniBooNE,
and other negative results.
Eq.\,(\ref{me4v2}) indicates that
the allowed region of LSND or MiniBooNE
should be the left side of the curve
$(1/4)\sin^22\theta_{ee}\sin^22\theta_{\mu\mu}$
but the MiniBooNE region is on the right side
of the ``disappearance(n)'' curve for all the values
of $\Delta m^2_{41}$ and the LSND region is
either on the right side the ``disappearance(n)'' curve
or it is disfavored by the negative results of other
experiments at 90\%CL.
Thus the results by LSND and MiniBooNE
are disfavored by the negative results of other
experiments.
On the other hand, if we take the affirmative results by
Neutrino-4 of $\nu_e\to\nu_e$
and IceCube of $\nu_\mu\to\nu_\mu$ seriously,
then the allowed region is the region which
is referred to as ``disappearance(a)'' in Fig.\,\ref{fig7}.
In this case, 
the mixing $\sin^22\theta_{\mu e}$ in the appearance probability
$P(\nu_\mu\to\nu_e)$ can be smaller
than the present upper bound
by more than one order of magnitude for
$\Delta m^2_{41}\sim$ 7 eV$^2$ or 9 eV$^2$,
and some part of the ``disappearance(a)'' region in Fig.\,\ref{fig7}
is still consistent with all other experiments in the past
except LSND and MiniBooNE.

The anomalies of LSND, reactor and gallium provide the main motivation
to study sterile neutrino oscillations.  Search for sterile neutrinos
in the same channel $\bar\nu_\mu\to\bar\nu_e$ as that of LSND
is still on-going\,\cite{JSNS2:2013jdh}, and it is hoped that these
anomalies will be confirmed in the future.

\subsection{Non-standard interactions in propagation}

If there is a 
flavor-dependent neutrino non-standard interaction (NSI) in neutrino
propagation:
\begin{eqnarray}
{\cal L}_{\mbox{\tiny{\rm NSI}}} 
=-2\sqrt{2}\, \epsilon_{\alpha\beta}^{ff'P} G_F
\left(\overline{\nu}_{\alpha L} \gamma_\mu \nu_{\beta L}\right)\,
\left(\overline{f}_P \gamma^\mu f_P'\right)\,,
\end{eqnarray}
where $f_P$ and $f_P'$ are the fermions with chirality $P$,
$\epsilon_{\alpha\beta}^{ff'P}$ is a dimensionless constant
normalized in terms of the Fermi coupling constant $G_F$,
then the matter potential in the flavor basis
is modified as
\begin{eqnarray}
A\left(
\begin{array}{ccc}
1+ \epsilon_{ee} & \epsilon_{e\mu} & \epsilon_{e\tau}\\
\epsilon_{\mu e} & \epsilon_{\mu\mu} & \epsilon_{\mu\tau}\\
\epsilon_{\tau e} & \epsilon_{\tau\mu} & \epsilon_{\tau\tau}
\end{array}
\right),
\label{matter-np}
\end{eqnarray}
where $A\equiv\sqrt{2}G_FN_e$
stands for the matter effect,
$\epsilon_{\alpha\beta}$ is defined by
$  \epsilon_{\alpha\beta}\equiv\sum_{f=e,u,d}
(N_f/N_e)\epsilon_{\alpha\beta}^{f}$
and $N_f~(f=e, u, d)$ stands for number densities of fermions $f$.
Here we defined the new NSI parameters as
$\epsilon_{\alpha\beta}^{f}\equiv\epsilon_{\alpha\beta}^{ffL}+\epsilon_{\alpha\beta}^{ffR}$
since the matter effect is sensitive only to the coherent scattering
and only to the vector part in the interaction.

It was pointed out\,\cite{Super-Kamiokande:2016yck}
that there was tension at 2$\sigma$ between the mass squared difference
from the solar neutrino data 
($\Delta m^2_{21}\simeq 4\times 10^{-5}$eV$^2$)
and that of the KamLAND experiment ($\Delta m^2_{21}=7.5\times 10^{-5}$eV$^2$).
Ref.\,\cite{Gonzalez-Garcia:2013usa}
showed that a nonvanishing value of the new NSI parameters
$\epsilon_{\alpha\beta}^f$ solves this tension.
This fact gives a motivation to take seriously NSI in propagation.
The significance
of the anomaly was approximately 2$\sigma$
in 2016\,\cite{Super-Kamiokande:2016yck}, but 
it has reduced to 1.4$\sigma$ in 2020\,\cite{SK:nu2020}.
If the significance keeps decreasing in the future, then
this should not be listed in the list of anomalies.

\section{Impact of Professor Koshiba on neutrino physics}

Today we have a lot of information on neutrinos, such as the two mass
squared differences, three mixing angles, and a hint of a CP violating
phase.  In the process of gaining such information, Professor
Masatoshi Koshiba made outstanding contributions to the field, by the
Kamiokande experiment which was proposed by him in
1979\,\cite{Watanabe:1979wu} and was started in 1983, and by the
Superkamiokande experiment which was again proposed by him in
1983\,\cite{Koshiba:1986jb}\footnote{It was initially named JACK
(Japan-America Collaboration at Kamioka), and was named
Superkamiokande in 1984\,\cite{Koshiba:1984rh}.} and was started in
1996.  The significant feature of the Kamiokande and Superkamiokande
experiments is that they give information on the energy spectrum and
the zenith angle dependence of neutrinos, which plays a crucial role
in discovering neutrino oscillations.  He even proposed an idea of a
Mega-ton class water Cherenkov detector in
1992\,\cite{Koshiba:1992yb}\footnote{ It was named DOUGHNUTS (Detector
Of Under-Ground Hideous Neutrinos from Universe and from Terrestrial
Sources) in Ref.\,\cite{Koshiba:1992yb}.}, and it is now being turned
into reality as the Hyperkamiokande experiment\,\cite{Abe:2011ts}.
Now it is evident that huge numbers of neutrino events are required to
determine the leptonic CP phase, and a Mega-ton class water Cherenkov
detector is ideal for that purpose.

In the standard framework of
three massive neutrinos,
the remaining quantities to be measured
are the sign of $\Delta m_{31}^2$, 
the sign of $\theta_{23}-45^\circ$, and
the CP phase $\delta$.  These quantities
are expected to be determined by the huge underground
neutrino experiments in the near future.
On the other hand, to complete the picture
of the Standard Model with three massive neutrinos,
we need to exclude the anomalies mentioned in Section \ref{np}.
If we cannot exclude or confirm them
by the experiments in the near future,
then we may need a new type of experiments, which even Professor
Koshiba could not imagine.

\section*{Acknowledgments}
When I was an undergraduate student, I had an opportunity to
participate in an experiment under Prof. Koshiba's guidance.  Some
twenty years later when we organized NuFact04 Summer Institute at
Tokyo Metropolitan University, Prof. Koshiba kindly accepted our
invitation to give a talk\,\cite{Koshiba:2004xx} for students and
postdocs, who were strongly impressed by his presentation.  I would
like to thank Prof. Koshiba for many things; for his guidance during
my undergraduate studies, for giving a wonderful talk at NuFact04
Summer Institute, and among others, for his great contributions to
neutrino physics.  This research was partly supported by a
Grant-in-Aid for Scientific Research of the Ministry of Education,
Science and Culture, under Grants No. 18K03653, No. 18H05543 and
No. 21K03578.

\end{document}